\documentclass[twocolumn,aps,prc,showpacs,preprintnumbers,amsmath,amssymb,10pt]{revtex4-1}

\usepackage{graphicx}
\usepackage{dcolumn}
\usepackage{setspace}
\usepackage{bm}
\usepackage{multirow}
\usepackage[dvipsnames]{xcolor}

\begin{document}
\title{Proton and neutron exchange as a prelude to fusion at near-barrier energies}

\author{J.~E. Johnstone}
\author{Varinderjit Singh}
\author{R. Giri}
\author{S. Hudan}
\author{J. Vadas}
\author{R.~T. deSouza}
\email{desouza@indiana.edu}
\affiliation{%
Department of Chemistry and Center for Exploration of Energy and Matter, Indiana University\\
2401 Milo B. Sampson Lane, Bloomington, Indiana 47408, USA}%

\author{D. Ackermann}
\author{A. Chbihi}
\author{Q. Hourdille}
\affiliation{
GANIL, CEA/DRF-CNRS/IN2P3, Bvd. Henri Becquerel, F-14076 Caen CEDEX, France}%

\author{A. Abbott}
\author{C. Balhoff}
\author{A. Hannaman}
\author{A.~B. McIntosh}
\author{M. Sorensen}
\author{Z. Tobin}
\author{A. Wakhle}
\author{S.~J. Yennello}
\affiliation{Cyclotron Institute and Department of Chemistry,
Texas A\&M University, College Station, Texas 77843 USA}

\author{M.~A. Famiano}
\affiliation{%
Department of Physics
Western Michigan University, Kalamazoo, Michigan 49008, USA}%

\author{K.~W. Brown}
\author{C. Santamaria}
\affiliation{%
National Superconducting Cyclotron Laboratory,
Michigan State University,
East Lansing, Michigan 48824}%

\author{J. Lubian}
\author{H.~O. Soler}
\affiliation{%
Instituto de F\'{i}sica, Universidade Federal Fluminense,
Niterói 24210-340, R.J., Brazil}%

\author{B.~V. Carlson}
\affiliation{%
Departamento de Fisica, Technologico da Aeronautica, Centro Tecnico Aerospacial,
12228-900 Sao Jose dos Campos, S\~{a}o Paulo, Brazil}

\date{\today}

\begin{abstract}
   Systematic examination of fusion for $^{39,41,45,47}$K + $^{28}$Si and $^{36,44}$Ar + $^{28}$Si provides insight into the impact of neutron and proton exchange on fusion for nuclei at and near the N=20 and N=28 shells. Comparison of the reduced excitation functions reveals a marked difference between the behavior of open-shell and closed-shell systems. While coupled channels calculations provide a good description for the closed-shell nuclei they significantly under-predict the fusion cross-section for open-shell nuclei. The observed trends are examined in the context of a potential energy surface, including shell effects, and multi-nucleon exchange with consideration of Pauli-blocking. 

\end{abstract}

 \pacs{21.60.Jz, 26.60.Gj, 25.60.Pj, 25.70.Jj}

\maketitle

\section{Introduction}

Nuclear reactions of neutron-rich nuclei play a key role in nucleosynthesis both in astrophysical environments \cite{Arnould07} as well as terrestially in accelerator-based experiments \cite{Oganessian10, Horowitz19}.
One topic of particular interest both theoretically as well as experimentally is the question about the enhancement or suppression of the fusion cross-section for neutron-rich nuclei \cite{Montagnoli17,KGA16, Canto09, Canto15}. For extremely neutron-rich nuclei, as a result of their weakly bound valence neutrons, one might observe reduced spatial coupling of the neutron and proton distributions and the emergence of novel neutron dynamics which enhance the fusion cross-section. 
At energies near the fusion barrier the fusion process is particularly interesting as the timescale of the collision is sufficiently long for collective dynamics of the neutron and proton density distributions to influence the fusion process. It is presently unclear how this dynamics is impacted by the shell structure of the initial nuclei.
Although it is well established that inelastic excitation of the two nuclei as they approach \cite{Stefanini95} and transfer of one or more nucleons \cite{Rowley92, Jia14} can modify the fusion probability in particular systems, a more comprehensive understanding is presently lacking \cite{Jia14}.

Theoretical calculations of the fusion using a density-constrained TDHF approach found an enhancement of fusion for the asymmetric system $^{24}$O + $^{16}$O as compared to $^{16}$O + $^{16}$O \cite{Umar12}. This enhancement is understood as resulting from neutron transfer which modifies the potential between the nuclei, lowering the barrier. For even more neutron-rich nuclei--at the limit of stability-- namely $^{24}$O + $^{24}$O, fusion is suppressed relative to $^{24}$O + $^{16}$O. This suppression of fusion for symmetric neutron-rich systems has been attributed to a repulsive Pauli potential 
arising from the overlap of the neutron-rich tails \cite{Simenel17}. 
However, these calculations are one-body and neglect many-body correlations which could enhance correlated transfer. Moreover, they are limited in that they only reflect the average behavior of the system.

Experimental evidence of fusion enhancement for neutron-rich nuclei also exists. Neutron exchange of valence neutrons in Ni + Ni systems were proposed as possibly responsible for an observed increase in the sub-barrier fusion cross-section \cite{Beckerman80}. Recent measurements provide further evidence of fusion enhancement due to the presence of a one-neutron halo ($^{15}$C) \cite{Alcorta11} or an unpaired neutron ($^{19}$O) \cite{Singh17}.  However, experimental measurements confirm that the neutron-richness of the colliding nuclei alone is not the only factor impacting the fusion probability as indicated by examination of Ca+Ca collisions. While fusion of a $^{48}$Ca projectile with a $^{40}$Ca target nucleus is enhanced as compared to a $^{40}$Ca projectile \cite{Aljuwair84}, fusion of $^{48}$Ca + $^{48}$Ca is suppressed below the barrier \cite{Montagnoli12}. It has recently been observed that, at above-barrier energies, after accounting for systematic size and Coulomb effects, the fusion cross-section for open shell nuclei near the N=20 and N=28 shells is larger than that of the closed-shell nuclei \cite{Singh21}. This result has been interpreted as enhanced binding of the closed-shell nuclei as compared to open-shell nuclei as they merge. 

In the present work, motivated by these prior above-barrier results, we examine for the first time fusion in $^{39,41,45,47}$K and $^{36,44}$Ar +$^{28}$Si and explore the role of shell structure and N/Z equilibration on the fusion cross-section.

\section{Experimental Data}

Radioactive beams of K and Ar ions were produced by the coupled cyclotron facility at MSU-NSCL and thermalized in a linear gas stopper before being re-accelerated by the ReA3 linac \cite{Singh21}. The re-accelerated beam was
transported to the experimental setup where it impinged upon the $^{28}$Si target. Details on the experimental setup have been previously published \cite{Vadas18}.

Contaminants in the radioactive beam were identified and rejected on a particle-by-particle basis by performing a $\Delta$E-TOF measurement \cite{Vadas18, Singh21}. 
The target composition was characterized using Rutherford
Backscattering measurement (RBS)  and confirmed using X-ray Photoelectron spectroscopy \cite{Johnstone20}. This RBS measurement revealed a $^{28}$Si thickness of 258 $\pm$ 10 $\mu$g/cm$^2$ and an oxygen thickness of 98  $\pm$ 4 $\mu$g/cm$^2$. The experimental resolution allowed reaction products from the fusion of the beam with $^{28}$Si and $^{16}$O to be distinguished \cite{Vadas18}. 
The intensities of the K and Ar beams incident on the target ranged between 1.0 x 10$^4$ ($^{44}$Ar/s) and 4.5 x 10$^{4}$ ($^{39}$K/s). Fusion of the incident K and Ar ions with the oxygen nuclei has been previously published \cite{Singh21}.

Fusion of K (Ar) ions
with the $^{28}$Si target results in a compound nucleus (CN) of As (Ge). 
De-excitation of the CN {\em via} neutron, proton and $\alpha$ emission deflects the resulting evaporation residue (ER) from the beam direction. The ER was detected
in annular Si(IP) detectors (1.0$^\circ$ $<$ $\theta_{lab}$ $<$ 7.3$^\circ$) and distinguished from scattered beam using the energy/time-of-flight (ETOF) technique \cite{Vadas18}. 

Extraction of the fusion cross-section, $\sigma_F$, is achieved by measuring the yield of ERs and utilizing the relation
$\sigma_F$ = N$_{ER}$/($\epsilon_{ER}$ $\times$ t $\times$ N$_{I}$) where 
N$_{ER}$ 
is the number of evaporation residues detected, N$_{I}$ is the number of beam particles of a given type incident on the target, t is the target 
thickness, and $\epsilon_{ER}$ is the detection efficiency. The number of detected residues, 
N$_{ER}$, is determined by summing the number of detected residues identified by the ETOF technique. Uncertainty in identifying an ER associated with fusion on
$^{28}$Si is reflected in the error bars presented.  
Beam particles with the appropriate identification in the $\Delta$E-TOF map provided the measure of
 N$_{I}$. 
A statistical model was 
employed to describe the de-excitation of the fusion product.  Together with the geometric acceptance of the experimental setup
this provided the detection efficiency, $\epsilon_{ER}$. which varied between $\approx$ 78-84\% over the 
entire energy range. 

An effective means of comparing the fusion excitation function for different systems is the use of the reduced excitation function \cite{Canto15}. Comparison of fusion for an isotopic chain allows utilization of the simplest scaling prescription.
The systematic increase in size with increasing mass number A is accounted for by scaling
the fusion cross-section $\sigma_F$ by the quantity (A$_P^{1/3}$  + A$_T^{1/3}$)$^2$. Differences in the Coulomb barrier for the different systems are considered for by examining the dependence of this reduced cross-section 
on the incident energy relative to the Coulomb barrier.
The Coulomb barrier, V$_C$, is taken as V$_C$=1.44Z$_P$Z$_T$/(1.4(A$_P^{1/3}$  + A$_T^{1/3}$)).  This simple accounting of the Coulomb barrier suffices as significant interpenetration of the charge distribution does not occur outside the fusion barrier. 

\begin{figure}
  \includegraphics[scale=0.52]{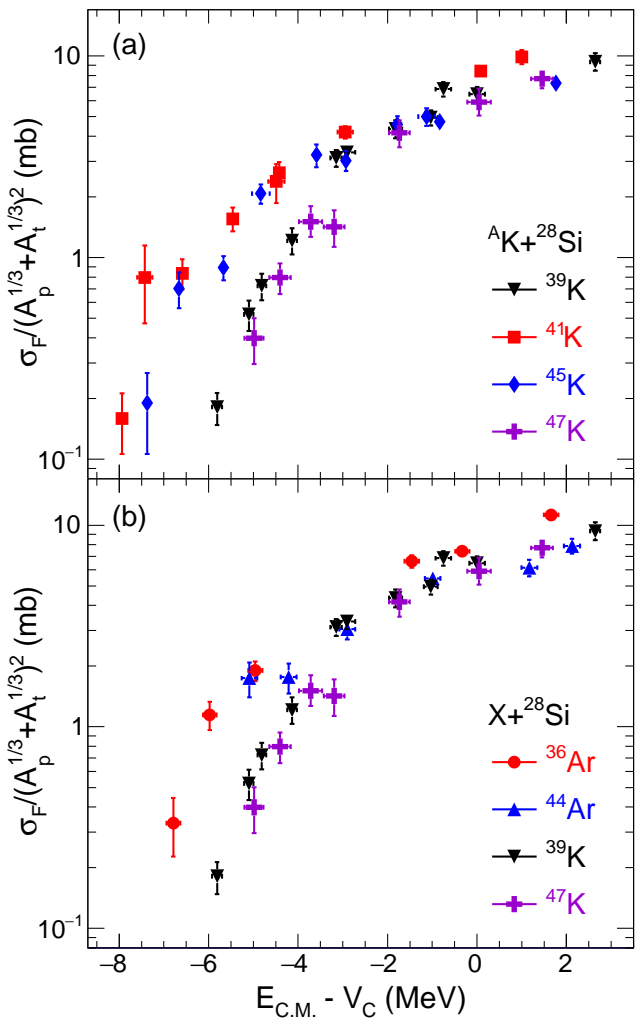}
  \caption{Comparison of the reduced  fusion excitation functions.
  }
  \label{fig:Fus_Red_VC}
\end{figure}

Presented in Fig.~\ref{fig:Fus_Red_VC}a are the reduced fusion excitation functions for $^{39,41,45,47}$K + $^{28}$Si. For all systems the reduced fusion cross-section above the barrier is similar. Below the barrier however, significant differences are apparent between the different systems. The data clearly organize into two groups: one associated with $^{39}$K and $^{47}$K (closed neutron shells at N=20 and N=28 respectively)
and the other with $^{41}$K and $^{45}$K (open neutron shells).
This similarity of the reduced fusion cross-section for $^{39}$K and $^{47}$K projectiles indicates that the density distributions, relevant to fusion, for the two closed-shell K isotopes are similar when scaled by A$^{1/3}$.
In marked contrast, 
a larger reduced fusion cross-section is evident for the open-shell $^{41}$K (N=22) and $^{45}$K (N=26), beyond the systematic A$^{1/3}$ scaling. 
This enhancement of the fusion cross-section for the open-shell nuclei increases 
with decreasing energy below the barrier. 
The same enhancement at sub-barrier energies is observed for the open-shell
$^{36,44}$Ar nuclei as compared to the closed-shell K isotopes  in Fig.~\ref{fig:Fus_Red_VC}b.
This present observation of the difference in the fusion of open-shell and closed-shell was 
confirmed by reexamining the literature. Enhancement of the fusion cross-section for an open-shell nucleus ($^{124}$Sn) as compared to a closed-shell nucleus ($^{132}$Sn) is also evident for $^{40,48}$Ca targets \cite{Kohley13}.

\begin{figure}
\includegraphics[scale=0.52]{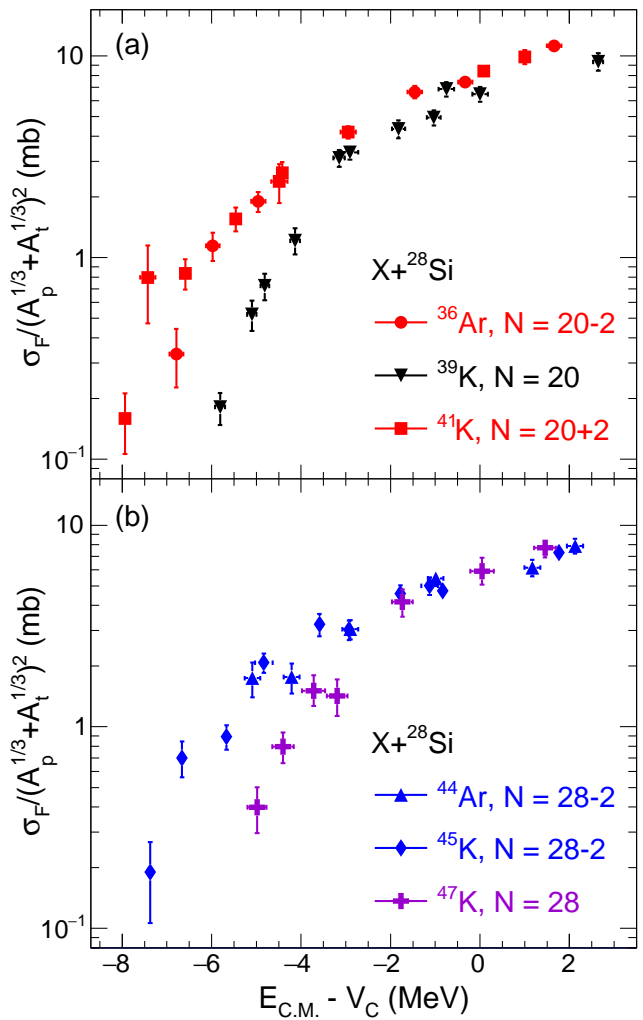}
\caption{ 
Reduced excitation functions for the measured systems. The upper panel shows
the systems closest to the N=20 shell closure.  The lower panel shows the systems
closest to the N=28 shell closure.
}
\label{fig:Fus_Red_VC_N}
\end{figure}

Presented in Fig.~\ref{fig:Fus_Red_VC_N} are the reduced fusion excitation functions grouped by their proximity to the N=20 and N=28 shells. In Fig.~\ref{fig:Fus_Red_VC_N}a one observes that $^{36}$Ar and $^{41}$K exhibit similar excitation functions with a marked enhancement of the reduced fusion cross-section as compared to the closed-shell $^{39}$K (N=20). This result indicates that the presence of two holes below the closed-shell ($^{36}$Ar) is effectively the same as the presence of two particles above the closed-shell ($^{41}$K) in determining the reduced fusion cross-section.
A similar enhancement in the reduced fusion cross-section is observed at the N=28 shell for the presence of two holes in $^{44}$Ar and $^{45}$K as compared to $^{47}$K.

\section{Comparison with Theoretical Models}

The simplest description of fusion involves the
interaction of the density distributions of the two interacting nuclei. For a non-adiabatic interaction (sudden approximation) consideration of the ground-state density distributions suffices. For adiabatic collisions, collective modes
in the colliding nuclei can be excited and also need to be considered. Inclusion of
these modes in a coupled channels (CC) formalism results in an increase in the
fusion cross-section at energies near and below the Coulomb barrier \cite{StG81,ADT89}.
To investigate whether the observed fusion excitation functions can
be described by the interaction of the density distributions of
the projectile and target nuclei, the S\~{a}o Paulo model was used.
The S\~ao Paulo potential (SPP)~\cite{CPH97} is a local equivalent double folding of the projectile and target matter densities on the zero-range interaction. 

Prior work demonstrated the sensitivity of the fusion cross-section to accurate ground-state density distributions \cite{Singh21}. To provide reasonably accurate matter density distributions, which include two-body correlations, we performed Dirac-Hartree-Bogoliubov (DHB) calculations \cite{CaH00}. The correlations in the DHB calculations of the present work are limited to surface-pairing correlations. These correlations can make
subtle modifications to the nuclear surface, extending and
modifying the nuclear density. The details of these mean field calculations using an axially-symmetric self-consistent approximation are reported in Ref.~\cite{CCG21}.

\begin{figure}
\includegraphics[scale=0.45]{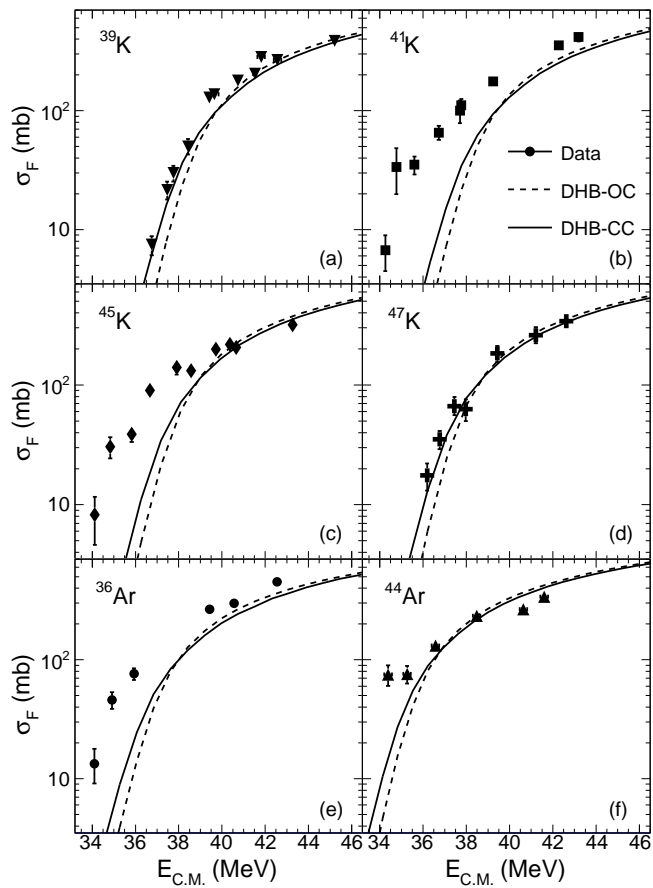}
\caption{ 
Comparison of the experimental cross-sections with the predictions of the S\~ao Paulo 
model using DHB densities for both the ground state and coupled channel calculations.}
\label{fig:Expt_vs_theory}
\end{figure}

Using the ground-state DHB matter distributions for both the projectile and $^{28}$Si target nuclei,
the SPP was generated and used to calculate the fusion cross-section.
The theoretical predictions, represented by the dashed lines, are compared with the experimental data in
Fig. ~\ref{fig:Expt_vs_theory}.  
Comparison of these one-channel (DHB-OC) theoretical predictions with the
experimental excitation functions is revealing.
For the closed neutron shell isotopes $^{39,47}$K, the DHB-OC calculations
provide a reasonable prediction of the excitation function over the entire energy interval measured although the model calculations lie slightly below the experimental data particularly in the sub-barrier regime.
In the case of the open neutron shell $^{41}$K,
$^{45}$K and $^{36}$Ar, the model dramatically under-predicts the measured cross-sections, particularly at sub-barrier energies. This
under-prediction for the case of the open-shell nuclei suggests that the ground-state configurations alone are insufficient in describing the measured cross-sections. In the case of $^{44}$Ar insufficient data exists at low energy to draw a definitive conclusion.

As coupling to low-lying collective modes acts to increase the fusion cross-section we have performed coupled-channels (CC) calculations to investigate the extent to which the presence of low-lying states increases the fusion cross-section. The 1.779 MeV, $2^{+}$ and 4.618 MeV, $4^{+}$ first states of the target were considered. The coupling to the low-lying projectiles states does not produce a considerable effect on the fusion cross-section. To account for the couplings between the low-lying states the transition probabilities were taken from Ref.~\cite{RNT01}.

The results of the CC calculations are shown in 
Fig. ~\ref{fig:Expt_vs_theory} as solid lines. In the case of the closed-shell nuclei, $^{39}$K and $^{47}$K, inclusion of the excitations considered provides a good description of the fusion cross-section. However, in the case of the open-shell nuclei the experimental data are significantly enhanced relative to the CC calculations with inclusion of low-lying excitations. It is particularly interesting to note that the magnitude of the enhancement is much larger than the increase due to the inclusion of inelastic excitation in the CC calculations. This enhancement suggests that transfer might be occurring prior to fusion.

\begin{table} [t!]

  \begin{center}
    \begin{tabular}{ccccc}
    \hline
    {Q} & {$^{39}$K} & {$^{41}$K} & {$^{45}$K} & {$^{47}$K}  \\ [0,4ex]
    \hline
    1n & -4.604 & -1.621 & -0.432 & 0.105 \\
    2n & -6.068 & 1.187 & 2.899 & 3.843 \\ 
    \end{tabular}
  \end{center}
  \caption{Q-values (in MeV) for one- and two-neutron transfer.}
  \label{table:Qval} 
\end{table}

Neutron transfer prior to fusion is often proposed as responsible for an enhancement in the fusion cross-section \cite{Beckerman80,Montagnoli13,Trotta01}. 
For a system with zero Q-value for two neutron transfer, $^{60}$Ni + $^{58}$Ni, inelastic excitations dominate and neutron transfer plays a negligible role \cite{Stefanini95}. When one of the colliding nuclei is neutron-rich relative to its collision partner, as in the case of $^{40}$Ca + $^{96}$Zr  positive Q-value neutron transfer channels act to increase the fusion cross-section at sub-barrier energies as compared to $^{40}$Ca + $^{90}$Zr \cite{Timmers97,Timmers98,Stefanini06,Stefanini14}. We present the relevant Q-values in Table~\ref{table:Qval} \cite{nudat}.
With the exception of $^{39}$K the Q-value for two-neutron transfer in the other K isotopes is positive.
Transfer of one neutron from $^{39}$K to $^{28}$Si is -4.604 MeV, while for $^{47}$K it is slightly positive (+0.1 MeV).  Nonetheless, the fusion excitation function for these two nuclei with $^{28}$Si is comparable. 
The Q-value for neutron transfer for the open-shell cases $^{41,45}$K lies between that of $^{39}$K and $^{47}$K yet the fusion excitation functions of the open-shell cases differ from those of the closed-shell. Clearly the observed 
behavior of the experimental fusion excitation functions cannot be understood simply by consideration of the Q-value for neutron transfer. 

The consideration of the Q-value for neutron transfer ignores the role of protons during the fusion process.  
Description of fusion using a density-constrained time-dependent Hartree-Fock (DC-TDHF) model allows the neutron and proton density distributions to evolve as the collision proceeds while incorporating all of the dynamical
entrance channel effects such as neck formation, particle
exchange, internal excitations, and deformation effects \cite{Oberacker13}. Such calculations for the system $^{132}$Sn + $^{40,48}$Ca clearly indicate the correlated flow of neutrons and protons. Unfortunately, for nuclei with unpaired nucleons the DC-TDHF calculations are considerably more challenging with a significant sensitivity to the inclusion of pairing \cite{Steinbach14a}.

\begin{figure}
\includegraphics[scale=0.43]{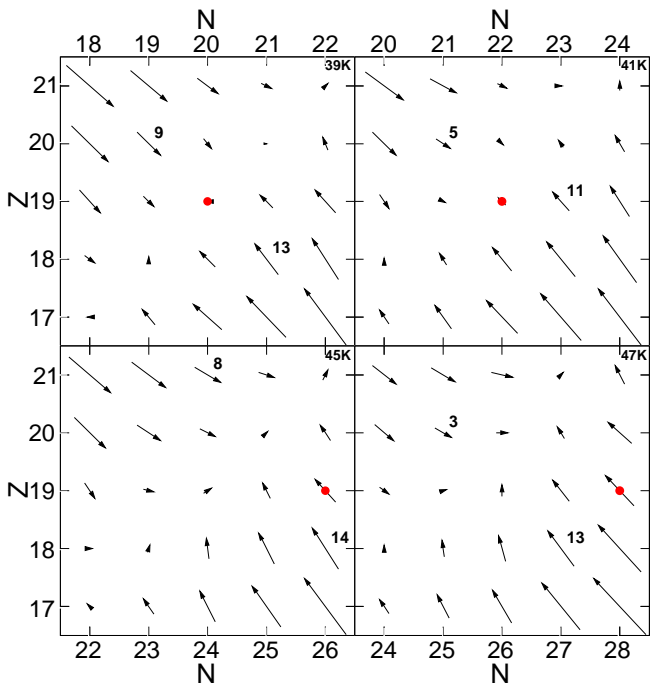}
\caption{ 
Potential energy surfaces (in MeV) for binary fragmentation of each of the K + $^{28}$Si systems. The arrows indicate the gradient of the liquid drop surface calculated with shell and proximity corrections for impact parameter zero. The initial projectile-target combination is indicated by the solid (red) symbol.
}
\label{fig:PES}
\end{figure}

We therefore consider qualitatively how initial nucleon exchange could impact fusion at near and sub-barrier energies using a conceptually simple physical framework. When the two colliding nuclei are within the range of the strong force nucleon exchange is allowed. This exchange of protons and neutrons is governed by a potential energy surface (PES). Flow of nucleons between the two nuclei is stochastic and allows equilibration of mass, charge, and energy \cite{Randrup79}.  The differential flow of neutrons and protons between the colliding nuclei results in both a net change in the atomic and mass numbers as well as excitation of the system. The nucleon flow is mitigated by Pauli-blocking of scattering into occupied states. 
Independent of the gradient of the potential, proton exchange is initially disfavored relative to neutron exchange because of the Coulomb barrier between the two nuclei. 
This physical picture was largely successful in explaining the charge and mass distributions associated with strongly damped collisions along with the characteristic dissipation of kinetic energy \cite{Schroder84}. 
A key factor driving the equilibration of N/Z in strongly-damped heavy-ion collisions is the gradient of the PES in the vicinity of the entrance channel \cite{desouza88}. A stochastic mean field approach utilizing this nucleon exchange framework successfully explained the dispersion of the mass distribution in $^{58}$Ni + $^{60}$Ni for damped collisions \cite{Yilmaz18}. It was hypothesized that for slightly more central collisions that resulted in fusion such a physical picture should still be valid. Unfortunately, 
the diffusion approach employed does not allow a description of the transition from multi-nucleon transfer to fusion \cite{Yilmaz18}.  We emphasize that in the present work we only utilize this physical picture to understand the factors influencing the {\em initial} neutron and proton exchanges prior to fusion.

To assess the factors influencing the initial nucleon exchanges, the PES was calculated for all binary combinations of a colliding system.  The PES calculated corresponds to the liquid drop energy modified by shell corrections as well as a proximity interaction \cite{Schroder84}. The surface was calculated at the strong absorption radius (approximately 10 fm in all cases shown) for zero impact parameter. As our aim is a qualitative description for these near and sub-barrier collisions and the systems considered are similar in mass asymmetry, ignoring the role of angular momentum in modifying the surface is justified.

The PES for each of the four K + $^{28}$Si systems is displayed in Fig. ~\ref{fig:PES}. Arrows indicate the gradient of the potential in the NZ plane with the initial projectile-target combination indicated by the solid (red) symbol. The magnitude of the gradient is indicated by the numbers (in MeV) adjacent to selected arrows. 

To begin we examine the cases of extremes in neutron-richness which nonetheless exhibit the same reduced fusion excitation function. In the case of $^{39}$K, the projectile-target combination already lies along the valley of the PES in the NZ plane. Therefore, correlated neutron and proton exchange is required in order to maintain N/Z equilibrium. While any initial proton transfer is disfavored because of the Coulomb barrier, proton transfer from $^{28}$Si to $^{39}$K is additionally suppressed by Pauli blocking \cite{Schroder84}. This suppression of initial proton exchange suppresses the neutron exchange. 

In the case of $^{47}$K, the PES is quite different. The initial system has a significant gradient to decrease the neutron number and increase the atomic number of the $^{47}K$ nucleus.  While neutron transfer out of the K nucleus is favored, proton pickup from the $^{28}$Si is also favored  due to the large N/Z asymmetry of the system. Pauli blocking of initial proton transfer limits the ability of the system to follow the gradient of the PES and attain N/Z equilibrium in an effective manner.

For the open neutron shell nuclei, neutron transfer is not hindered by the energy cost of breaking the neutron shell.
For $^{41}$K, as indicated by the PES, transfer of a neutron from $^{41}$K to $^{28}$Si can  occur without any driving force for proton transfer. Net transfer of one neutron in this physical picture corresponds to multiple neutron exchanges. These multiple neutron exchanges excite the K nucleus which lessens the Pauli-blocking of subsequent proton exchanges. Subsequent proton transfer into or out of the K nucleus are equally energetically favorable as indicated by the PES. One might hypothesize that these initial neutron exchanges, not just the net transfer of one neutron, by reducing the Pauli-blocking act to increase the fusion probability.

The case of the $^{45}$K is intermediate between that of $^{41}$K and $^{47}$K and more difficult to interpret. While pickup of a proton by the $^{45}$K is favored along with loss of a neutron, the magnitude of the gradient is less than in the $^{47}$K case. The smaller driving force for proton pickup relative to $^{47}$K suggests a lesser role of proton transfer on the fusion cross-section.

\section{Conclusions}

Comparison of the fusion excitation functions for
$^{39,41,45,47}$K + $^{28}$Si and $^{36,44}$Ar + $^{28}$Si reveals that at sub-barrier energies the open neutron shell nuclei of
$^{41,45}$K manifest a significantly larger reduced fusion cross-section as compared to the closed neutron shell isotopes $^{39,47}$K. 

For the closed-shell nuclei, the
use of  Dirac-Hartree-Bogoliubov (DHB)
ground state densities in the S\~ao Paulo
fusion model provided a reasonable description of the data - one that was improved by inclusion of low-lying states of the $^{28}$Si. For the open-shell nuclei, use of the DHB densities, even with the inclusion of the excited states, significantly under-predicts the measured
cross-sections, particularly below the barrier. 
Q-value calculations of neutron transfer alone are unable to explain the similarity in cross-section for the closed-shell nuclei and the enhancement for the open-shell nuclei. 
If transfer is the reason for the enhancement, a slightly more expansive perspective is required.

Consideration of the energetics involved with {\em both} proton and neutron exchange, along with Pauli-blocking, provided insight into the difference between the closed-shell and open-shell nuclei.  A more quantitative description of the observations requires development of a more complete theoretical description, one which properly accounts for multi-nucleon transfer and Pauli-blocking in the initial stages of the collision.

\begin{acknowledgments}
We acknowledge the high quality beams provided by the staff at NSCL, Michigan Sate University that made this experiment possible.
This work was supported by the U.S. Department of Energy Office of Science under Grant Nos. 
DE-FG02-88ER-40404 (Indiana University), DE-FG02-93ER-40773 (Texas A\&M University) and the National Science Foundation under PHY-1712832. Brazilian authors acknowledge  partial  financial  support  from  CNPq,  FAPERJ, FAPESP, CAPES and INCT-FNA (Instituto Nacional de Ci\^ {e}ncia e Tecnologia- F\' isica Nuclear e Aplica\c {c}\~ {o}es)  Research Project No.  464898/2014-5

\end{acknowledgments}


%

\end{document}